\documentclass[12pt]{article}
\usepackage{graphicx}
\usepackage{latexsym}
\usepackage{mathptm}
\usepackage{amsmath}
\usepackage{amssymb}
\usepackage{graphicx}
\newcommand{\f}{\begin{equation}}
\newcommand{\ff}{\end{equation}}
\newcommand{\tr}{\text{Tr}}
\newcommand{\str}{\text{STr}}

\begin{document}

\title{An action for higher spin gauge theory in four dimensions}
\author{Nima Doroud\thanks{ndoroud@perimeterinstitute.ca} \ and Lee Smolin\thanks{lsmolin@perimeterinstitute.ca}\\
\\Perimeter Institute for Theoretical Physics, 
\\31 Caroline St. N. Waterloo, Ontario N2L 2Y5, Canada
\\ and
\\Department of Physics, University of Waterloo, 
\\200 University Ave. W. Waterloo, Ontario N2L 3G1, Canada}
\renewcommand{\today}{February 2011}
\maketitle

\begin{abstract}

An action principle is presented for Vasiliev's Bosonic higher spin gauge theory in four spacetime dimensions.  The action is of the form of a broken topological field theory, and arises by an extension of the MacDowell-Mansouri formulation of general relativity.  In the latter theory the local degrees of freedom of general relativity arise by breaking the gauge invariance of a topological theory from $sp(4)$ to the Lorentz algebra.  In Vasiliev's theory the infinite number of degrees of freedom with higher spins similarly arise by the breaking of a topological theory with an infinite dimensional gauge symmetry extending $sp(4)$ to the Lorentz algebra.  

The Hamiltonian formulation of Vasilev's theory is then derived from our action, and it is shown that the Hamiltonian is a linear combination of constraints, as expected for a diffeomorphism invariant theory. The constraint algebra is computed and found to be first class.  

\end{abstract}
\newpage

{ \footnotesize
\tableofcontents}

\newpage


\section{Introduction}

One of the most basic lessons of modern theoretical physics is the intimate connection between consistent dynamics of fields with spin and gauge invariance \cite{Plebanski, Capovilla, Mansouri, Stelle}. This is why the degrees of freedom of spin one and spin two fields are expressed as connections, whose dynamics is coded by action principles which are functionals of the corresponding curvatures.

When the theory is also diffeomorphism invariant, as in the case of general relativity and supergravity, there is a further bit of wisdom, which is that the dynamics of the theory is closely related to that of a topological field theory.  A topological field theory is one in which the field equations include the vanishing of a curvature, so that there are no local degrees of freedom in the bulk spacetime.  Degrees of freedom exist, but only on boundaries or associated with holonomies of non-contractible loops. 

It turns out that general relativity and supergravity in different dimensions can be understood as arising from such topological field theories by either symmetry breaking or the imposition on the field equations of constraints.   This was done in different ways by Pleba\'{n}ski \cite{Plebanski,Capovilla} and by MacDowell and Mansouri \cite{Mansouri, Stelle}. Extensions to different dimensions \cite{higher} and to supergravity theories in both four and eleven dimensions \cite{super,11d} have shown the power of this insight.

A very closely related insight is contained in an approach known as the unfolding formalism which makes use of the mathematical structures of free differential algebras.  
These can be seen as a kind of convergent evolution which captures the same structures as topological field theory in a slightly different langauge.   The language of topological field theory connects these insights to a rich mathematical setting related to topological invariants in three and four dimensions and quantum algebras and conformal field theory, while the language of free differential algebras is closely related to the structure of supergravity theories.  

Whichever language is used, these formulations have several advantages because the  action and field equations are polynomial. Indeed, the action can be made cubic so the field equations are quadratic, so that these gravitational theories are expressed in the least non-linear way possible. A related fact is that  
the diffeomorphism invariance can be understood as arising from local gauge symmetries.  By putting diffeomorphism invariant theories in the simplest possible
form, these formulations make it possible to see how the  delicate conditions needed to embed a small number of dynamical degrees of freedom within a 
complex of auxiliary fields and gauge degrees of freedom are accomplished consistently.  

In this letter, these insights are extended to a remarkable  theory developed over the last years by Vasiliev and collaborators, which is a diffeomorphism invariant theory describing the interactions of fields of arbitrarily high spin\footnote{As we were finishing this paper we learned that this problem has been addressed independently by Boulanger and Sundell \cite{Sundell}.} \cite{Fradkin1987, Vasiliev1989, Fradkin1987_2, Vasiliev1990, Vasiliev1996, Vasiliev2003, Bekaert2005}.  It is rather remarkable that a theory with interacting higher spin fields exists, but it does, as expressed in a set of beautiful field equations written down by Vasiliev.   A key open problem has been to construct an action for this theory.  By using the insights of the methods of broken topological field theories or, equivalently, the unfolded formalism, we give such a construction here.  

The construction we give here reveals Vasiliev's theory to be a natural extension of general relativity, when the latter is looked at as a broken topological field theory. In the Macdowell-Mansouri formulation, the dynamical fields are all coded into a connection, $A_a^{AB}$ valued in $sp(4)$, (locally isomorphic to $\mathfrak{so}(5)$\footnote{for simplicity we speak here of Euclidean signature.} )where $A=(0,1,\dot{0},\dot{1})$ are four dimensional spinor indices, which can be thought of as components of a Dirac spinor. One proceeds to write an action which is invariant under local $Sp(4)$ gauge transformations and then break it to the Lorentz group, $SO(4)$.  As we will review below, the frame fields of spacetime are expressed as the components of the connection in the coset $sp(4)/\mathfrak{so}(4)$ while the unbroken $\mathfrak{so}(4)$ part of the connection becomes the spin connection.  

Crucial to this way of understanding general relativity is that it only works when the cosmological constant, $\Lambda$, is non-vanishing.  
A further insight into the role of the cosmological constant was gained in a refinement of the Macdowell-Mansouri formulation \cite{artem} in which the $Sp(4)$
symmetry is broken only when the cosmological constant is non-zero. When $\Lambda =0$ the gauge symmetry is the full $Sp(4)$ and the theory reduces to a
topological field theory. 

The Vasiliev theory is constructed by an extension of this strategy in which the $sp(4)$ Lie algebra is extended to an infinite dimensional lie algebra called, $hs(4)$.  One constructs a connection $A_a$ valued in $hs(4)$ and its corresponding curvature, $F_{ab}$.  The Vasiliev field equations put constraints on the curvature which express a breaking to its local Lorentz subalgebra. This is closely analogous to the Macdowell-Mansouri formulation of general relativity which arises by field equations that break $sp(4)$ to $\mathfrak{so}(4)$.  As we show below, these field equations can be derived from an action principle which expresses the breaking of $hs(4)$ to $\mathfrak{so}(4)$.   Indeed, as in the Macdowell-Mansouri formulation, Vasiliev's theory does not have a consistent interacting limit as $\Lambda\rightarrow 0$.

\subsection{Background}

To make this paper accessible to readers who haven't been introduced to  higher spin gauge theories, we give here a brief history of their study. While the non-interacting theory of higher spins had been well studied\footnote{see for example \cite{freehs} and references therein.} it was commonly believed that a consistent gravitational interaction among higher spin fields does not exist. This was partly due to the results of previous attempts to introduce gravitational interactions (in flat spacetime) \cite{hsintproblems} and partly due to the infamous Coleman-Mandula theorem \cite{ColeMan} and it's generalization by Haag \emph{et al.} \cite{Haag}. 

It was not until 1987 that the first results pointing towards a consistent interacting theory were obtained \cite{Fradkin1987, Vasiliev1989}. In their first paper on the subject, Vasiliev and Fradkin showed that the gravitational interaction of massless higher spin fields does indeed exist to the first nontrivial order. This result cast new light on the previous attempts as it clarified the crucial role played by the cosmological constant $\Lambda$, \emph{i.e.} the interactions were shown to be non-linear in $\Lambda$ and therefore admit no flat spacetime limit, $\Lambda\rightarrow 0$. This was followed by the identification of the appropriate class of gauge algebras \cite{Fradkin1987_2} which led to the formulation of the full interacting theory in 2+1 dimensions by Blencowe \cite{Blencowe1989} and in 3+1 dimensions, at the level of the equations of motion, by Vasiliev \cite{Vasiliev1990}. Vasiliev's equations of motion were later generalized to all dimensions \cite{Vasiliev1996,Vasiliev2003,Bekaert2005}. 

More recently, there has been a growing interest in higher spin gauge theories following the conjecture by Klebanov \& Polyakov \cite{Klebanov2002}, that Vasiliev's theory is AdS dual of the (critical) $O(N)$ vector model, for which there has since been a lot of evidence (see for example \cite{Giombi2010}). The asymptotic symmetries of the higher spin gauge theory in $2+1$ dimensions and the connection with $\mathcal{W}$-algebras have also been studied \cite{asymptotic}.

Despite having the consistent equations of motion at hand, there was  little progress in forming an action principle leading to Vasiliev's equations in 3+1 (and higher) dimensions. This is mainly a consequence of the abundance of auxiliary fields that need to be introduced in 4 and higher dimensions. The main result of this paper is a resolution of that problem, based on the methodology we described above for the simplest case, that is, massless Bosonic higher spin gauge fields in 3+1 dimensions with the gauge algebra $hs(4)$.

The plan of the paper is as follows; in the next section, the underlying mathematical structure of Vasiliev's higher spin gauge theory, namely, free differential algebras and unfolded formulation, is introduced. Section 3 is devoted to defining and re-deriving some of the properties of the higher spin gauge algebra, $hs(4)$, followed by the presentation of Vasiliev's theory in section 4. In section 5, we propose an action principle for Vasiliev's theory of massless Bosonic higher spin gauge fields in 4 dimensions. In section 6 we  carry out the constraint analysis of the theory and we close the paper with a summary of our results.

\section{Free differential algebras and unfolded formulation}

The great insight of Vasiliev that led to the formulation of consistent interacting higher-spin gauge theories was that a consistent set of equations can be written down for the dynamics of any set of fields via the so called ``\emph{unfolded formulation}". The general idea is to start with the fields on an appropriate principal fiber bundle equipped with a free (graded) differential algebra (FDA). One may then embed the equations of motion as a set of ``flatness" and covariant constancy conditions on the fields. The problem then reduces to finding the explicit form of the generalized curvatures of the FDA.

We begin this section with a brief introduction to free differential algebras which we use to sketch the general idea of unfolded formulation. An introduction to FDAs can be found in \cite{Castellani} while the application to unfolded formulation can be found in almost all papers on Vasiliev's theory, see for example \cite{Bekaert2005, Iazeolla}.

\subsection{Free differential algebras (FDA)}

Let $\mathcal{E}$ be a fiber bundle over $\mathcal{M}$, a $d$-dimensional manifold equipped with an exterior algebra $\Omega(\mathcal{M})$. Consider a set of differential forms $\{A^{i} : A^{i}\in\Omega^{p_{i}}(\mathcal{M})\}$ valued in the fiber. We may define the generalized curvatures $F^{i}$ as
\begin{eqnarray}
\label{FDAcurvature}
F^{i}=dA^{i}+G^{i}(\{A^{j}\}),
\end{eqnarray}
where $d$ denotes the exterior derivative operator satisfying $d^{2}\equiv 0$, and $G^{i}$ are $(p_{i}+1)$-\emph{forms} constructed
out of the wedge product of $A^{i}$, \emph{i.e.}
\begin{eqnarray}
\label{FDAG}
G^{i}=\sum_{n=1}^{\infty} g^{i}_{j_{1}\dots j_{n}}\delta_{_{p_{i}+1,p_{j_{1}}+\dots+p_{j_{n}}}}
A^{j_{1}}\wedge\dots\wedge A^{j_{n}}\ .
\end{eqnarray}
An FDA is defined by demanding the compatibility of $F^{i}=0$ and $d^{2}=0$. This imposes the \emph{generalized Jacobi identity} on $G^{i}$,
\begin{eqnarray}
\label{FDA}
G^{j}\wedge\frac{\delta G^{i}}{\delta A^{j}}=0\ ,
\end{eqnarray}
where implicit summation over $j$ is understood. Taking the exterior derivative of (\ref{FDAcurvature}) and using $d^{2}=0$ and the Jacobi identity (\ref{FDA}) we arrive at the  \emph{generalized Bianchi identity}
\begin{eqnarray}
\label{FDABianchi}
dF^{i}-F^{j}\wedge\frac{\delta G^{i}}{\delta A^{j}}=0\ .
\end{eqnarray}
An FDA may be viewed as a generalization of $G$-bundle strucutre. Indeed, for connection $1$-\emph{forms} valued in a Lie algebra $\mathfrak{g}$, an FDA is just a gauge theory, provided that the functions $G^{i}$ are quadratic. One may introduce generalized gauge transformations
\begin{eqnarray}
\label{FDAgauge}
\delta A^{i}=d\epsilon^{i}-\epsilon^{j}\wedge\frac{\delta G^{i}}{\delta A^{j}}\ ,
\end{eqnarray}
where $\epsilon^{i}$ is a $(p_{i}-1)$-\emph{form}. Evidently, $0$-\emph{form} connections do not give rise to any gauge parameters. The corresponding transformation of the generalized
curvatures is given by
\begin{eqnarray}
\label{FDAcurvegauge}
\delta F^{i}=-F^{k}\wedge\frac{\delta}{\delta A^{k}}\left(\epsilon^{j}
\frac{\delta G^{i}}{\delta A^{j}}\right)\ ,
\end{eqnarray}
so that the equations $F=0$ are gauge invariant. 

A fundamentally important remark is the following: whenever $F^{i}$ vanish, spacetime diffeomorphisms
\begin{eqnarray}
\label{diffeo}
\delta_{\xi}A^{i}=\mathcal{L}_{\xi}A^{i}=\{d,i_{\xi}\}A^{i}\ ,
\end{eqnarray}
reduce to gauge transformations with the specific choice of the gauge parameter $\epsilon^{i}=i_{\xi}A^{i}$.  This is the key to the role of topological
field theories, or FDA's in the classical and quantum dynamics of gravitational theories.  

\subsection{The unfolding strategy}

The unfolded formulation of a dynamical theory has the structure of an FDA with the equations of motion of the free theory given by
\begin{eqnarray}
\label{feq0}
F^{i}=0\ .
\end{eqnarray}
As it has been illustrated by Vasiliev and other authors\footnote{See for example \cite{Iazeolla}. Also, the unfolding of a free scalar field is studied in glorious detail in \cite{Shaynkman}.}, given the dynamical equations of a theory, one can always unfold them by virtue of introducing enough auxiliary fields (via addition of a fiber bundle structure). Note that the resulting unfolded theory, despite having the same on-shell dynamics, is not equivalent off-mass-shell to the standard theory that we started with.

Conversely, given a set of physical fields, one may form a suitable FDA structure such that equations (\ref{feq0}) define a consistent free theory describing the dynamics of the physical fields. To include interactions however, one has to deform the equations (\ref{feq0}) in a consistent manner which respects the symmetries of the theory. Let us illustrate the unfolding strategy in a few steps which are in direct analogy to the construction of gauge theories.

Consider a set of ``physical fields" $A^{i}$ which are $p_{i}$-\emph{forms} over a (4d) manifold $\mathcal{M}$, equipped with an exterior algebra $\Omega(\mathcal{M})$, which in particular include a $0$-\emph{form}\footnote{As we shall emphasize later on, this is necessary in order to incorporate non-trivial gravitational dynamics into the theory.}. We begin by adding a fiber bundle structure over $\mathcal{M}$ and promoting the physical fields to sections of the added fiber; in doing so, we automatically add a large number of auxiliary fields to the theory. By adding the fiber bundle structure, we make room for non-trivial extension of the exterior derivative of the form
\begin{eqnarray}
\label{dbar}
\bar{d}=d+\sigma\ ,
\end{eqnarray}
where $\sigma$ is a $1$-\emph{form} ``differential" operator which in general depends on the physical as well as auxiliary fields and does not involve spacetime derivatives. This may be viewed as a direct generalization of the connection in gauge theories. To realize the FDA structure, we rewrite $\sigma$ as
\begin{eqnarray}
\label{sigma}
\sigma=G^{j}\frac{\delta}{\delta A^{j}}\ .
\end{eqnarray}
Inserting this into (\ref{dbar}) and setting $d^{2}=0$, we deduce that $\bar{d}^{2}=0$ is equivalent to the set of equations (\ref{feq0}). The compatibility of $\bar{d}^{2}=0$ and $d^{2}=0$ will then impose the generalized Jacobi identity (\ref{FDA}) on $G^{i}$. 

To revive the standard gauge theory structure, we restrict the physical fields to a $1$-\emph{form}, $A$, and a $0$-\emph{form}, $C$, and the fiber to the product of a Lie algebra $\mathfrak{g}$ and a linear space $V$ carrying a representation $\rho$ of $\mathfrak{g}$ . The gauge potential $A$ lives in $\mathfrak{g}$ while the scalar $C$ lives in $V$. The generalized curvatures are then given by
\begin{eqnarray}
\label{FDAgencurv1}
F^{i}&=&dA^{i}+\frac{1}{2}g^{i}_{jk}A^{j}\wedge A^{k},\\
\label{FDAgencurv0}
R^{a}&=&dC^{a}+A^{i}\ \rho(t_{i})^{a}_{\ b}C^{b}\ ,
\end{eqnarray}
where $g^{i}_{jk}$ are the structure constants of $\mathfrak{g}$. The generalized Jacobi identities (\ref{FDA}) reduce to
\begin{eqnarray}
\label{genjacob1}
g^{i}_{[jk}g^{j}_{l]m}=0,\quad\text{and}\quad 
\rho(t_{[i|})^{a}_{\ b}\rho(t_{|j]})^{b}_{\ c}=\frac{1}{2}g^{k}_{ij} \rho(t_{k})^{a}_{\ c}\ ,
\end{eqnarray}
which are automatically satisfied. The vanishing of the curvatures (\ref{FDAgencurv1}) and (\ref{FDAgencurv0}) imply that the connection $A$ is flat and that $C$ is covariantly constant. By construction, the set of equations $F=0$ and $R\equiv DC=0$, where $D=d+A^{i}\rho(t_{i})$ denotes the covariant derivative, are gauge invariant and are compatible with $d^{2}=0$. To illustrate the details of the unfolding procedure, we end this section by showing how the MacDowell-Mansouri formulation of general relativity, can be understood as an example of an  unfolded formulation.

\subsection{The MacDowell-Mansouri formulation of general relativity}

Starting with the work of MacDowell and Mansouri \cite{Mansouri}, Stelle and West \cite{Stelle}, and further developments by others \cite{Plebanski, Capovilla,artem}, it is now well understood that we can reformulate the general theory of relativity\footnote{More precisely, on should speak of an extension of GR since this formulation admits degenerate solutions.}  (GR) in $d$ dimensions in terms of a gauge theory with $SO(2,d-1)$ gauge group. The corresponding action in 4 spacetime dimensions is of the form
\begin{eqnarray}
\label{MMSW}
S=\frac{1}{2G \Lambda}\int_{\mathcal{M}}\tr\big(\Gamma F\wedge F\big)-
\frac{1}{2G \Lambda}\int_{\partial\mathcal{M}}\tr\big(A\wedge dA+\frac{2}{3}A\wedge A\wedge A\big),
\end{eqnarray}
where $F=dA+A\wedge A$, is the curvature 2-\emph{form} corresponding to the $\mathfrak{so}(2,3)$ connection 
\begin{eqnarray}
\label{so23connection}
A = \frac{1}{2i}A^{BC}J_{BC}\ ,\quad B,C=0,\dots,3,5\ .
\end{eqnarray}
Here, $J_{AB}=\frac{{i}}{4}[\gamma_{A},\gamma_{B}]$ are the generators of the $AdS$ group with $\gamma_{A}$ denoting the Dirac matrices which span the Clifford algebra $C\ell_{2,3}$.

One may recognize (\ref{MMSW}) as a Topological action with an extra factor $\Gamma$ which is a $0$-\emph{form} valued in the Clifford algebra. As was mentioned earlier, this factor is necessary in order to have non-trivial dynamics. Indeed when $\Gamma=1$, the above action is topological and has no local degrees of freedom. The second term in the action (\ref{MMSW}) is the \emph{Chern-Simons form} which is the standard boundary term for the topological action. In the GR sector, the boundary term will play the role of the \emph{Gibbons-Hawking} boundary term which is to cancel the boundary variations of the bulk action. 

We constraint $\Gamma$ to be timelike and covariantly constant $D\Gamma=0$. We may then perform a partial gauge fixing which sets $\Gamma=\gamma^{5}$ and therefore breaks down the internal redundancy to $SO(1,3)$, the Lorentz group. We may then identify different components of the connection
\begin{eqnarray}
\label{so13connection}
A^{ab} &=& \omega^{ab},\\
\label{vierbein}
A^{a5} &=& \frac{1}{l} e^{a}\ ,
\end{eqnarray}
with the $\mathfrak{so}(1,3)$ connection, $\omega$, and the vierbein $e$. Here, $l$ is a constant of dimention length and is related to the cosmological constant via $\Lambda=-\frac{1}{l^{2}}$. Consequently, the curvature $F$ may be decomposed into
\begin{eqnarray}
\label{F23eqF13plusee}
F^{ab} &=& F_{\omega}^{ab}+\frac{1}{l^{2}}e^{a}\wedge e^{b},\\
\label{F23eqtorsion}
F^{a5} &=& \frac{1}{l}D_{\omega}e^{a},
\end{eqnarray}
with $F^{ab}_{\omega}$ and $D_{\omega}$ denoting the Lorentz curvature and covariant derivative, respectively. Inserting these back into the action and setting the torsion $F^{a5}$ to zero, the bulk action reduces to the well known Palatini action with a cosmological term.

Another way to write the MacDowell-Mansouri formulation is to introduce a set of auxiliary two forms $B$ valued in $sp(4)$ and write (\ref{MMSW}) as
\f
\label{bwedgef}
S= -\frac{l}{2G}\int_{\cal M}\Big[B^{AB} \wedge F_{AB} +\frac{1}{2} \epsilon_{ABCDE}  B^{AB} \wedge B^{CD }v^E - \mu (v^A v_A+l^{-2} )\Big]\ ,
\ff
where $v^E$ is a scalar field in a vector multiplet of $\mathfrak{so}(2,3)$. constrained by the variation of $\mu$ to lie on the four sphere \cite{artem}. Note that we can replace this constraint by a covariant constancy constraint $D_{A}v^{B}=0$, supplemented by the assumption that $v^{A}$ is everywhere timelike. The field equations are of the form
\f
F_{AB}= \epsilon_{ABCDE} B^{CD} v^E
\ff
characteristic of a broken or constrained topological field theory.  We will see shortly that the field equations of Vasiliev's theory take the same general form.

Before turning our attention to Vasiliev's theory, we would like to draw your attention to the following: the quantities $F_{ab}$, $F^{a5}$ and $D_{A}\Gamma$ (or equivalently, $D_{A}v^{B}$), may be viewed as the curvatures of an FDA and the vanishing of these curvatures describes Anti-de Sitter spacetime. The appropriate deformation of these curvatures that gives rise to general relativity can be obtained by varying either of the actions (\ref{MMSW}) or (\ref{bwedgef}).

\section{Higher spin gauge theory in 3+1 dimensions}

We now turn to Vasiliev's theory.  The key to understanding its structure is to extend $sp(4)$ in the construction of general relativity just sketched to an infinite
dimensional algebra called $hs(4)$. The aim of this section is to introduce this algebra\footnote{Here we only consider the simplest Lie algebra corresponding to the massless Bosonic higher spin fields. In 4 spacetime dimensions, this algebra may be constructed by starting from the $sp(4)$, or equivalently, $so(2,3)$, Lie algebra. While the case of the AdS algebra can easily be generalized to higher dimensions, one would have to deal with factoring out the ideals when taking the tensor product of fields living in the Lie algebra. Since we are only interested in the 4 dimensional theory 
we choose to start from the $sp(4)$ which is free of these complications.}. 

\subsection{The (Bosonic) higher spin Lie algebra $hs(4)$}

To begin with, we define the Weyl spinor doublets $y_{\alpha}$ and $\bar{y}_{\dot{\alpha}}$ $(\alpha,\dot{\alpha}=1,2)$, such that $(y_{\alpha})^{\dagger}=\bar{y}_{\dot{\alpha}}$. If we require these spinors to satisfy the commutation relations
\begin{eqnarray}
\label{ycommutation}
 [y_{\alpha},y_{\beta}]=2i\epsilon_{\alpha\beta}\ ,\
 [\bar{y}_{\dot{\alpha}},\bar{y}_{\dot{\beta}}]=2i\epsilon_{\dot{\alpha}\dot{\beta}}\ ,\ 
 [y_{\alpha},\bar{y}_{\dot{\beta}}]=0\ ,
\end{eqnarray}
then they form the oscillator realization of the $sp(4)$ algebra. The generators of $sp(4)$ are then given by
\begin{eqnarray}
\label{sp4generators}
 L_{\alpha\beta}=\frac{1}{2}\{y_{\alpha},y_{\beta}\}\ ,\ 
 \bar{L}_{\dot{\alpha}\dot{\beta}}=\frac{1}{2}\{\bar{y}_{\dot{\alpha}},\bar{y}_{\dot{\beta}}\}\ ,\
 P_{\alpha\dot{\beta}}=y_{\alpha}\bar{y}_{\dot{\beta}}\ .
\end{eqnarray}
One can easily verify using the commutation relations for the oscillators, equations (\ref{ycommutation}), that these generators satisfy the $sp(4)$ algebra,
\begin{eqnarray}
\label{commutation-sp(4)}
\left[L_{\alpha\beta},L_{\gamma\delta}\right]&=&2i\left(\epsilon_{\alpha\gamma}L^{}_{\beta\delta}+
\epsilon_{\alpha\delta}L_{\beta\gamma}+\epsilon_{\beta\delta}L_{\alpha\gamma}+
\epsilon_{\beta\gamma}L_{\alpha\delta}\right)\ ,\\ \nonumber &&\\
\left[L_{\alpha\beta},P_{\gamma\dot{\delta}}\right]&=&2i\left(\epsilon_{\alpha\gamma}
P_{\beta\dot{\delta}}+\epsilon_{\beta\gamma}P_{\alpha\dot{\delta}}\right)\ , \\ \nonumber &&\\
\left[P_{\alpha\dot{\beta}},P_{\gamma\dot{\delta}}\right]&=&2i\left(\epsilon_{\gamma\alpha}
L_{\dot{\beta}\dot{\delta}}+\epsilon_{\dot{\delta}\dot{\beta}}L_{\alpha\gamma}\right)\ ,
\end{eqnarray}
with similar relations for $\bar{L}$. The next step is to define the (Bosonic sector\footnote{One can include Fermionic fields by lifting the restriction $n+m=$even, in (\ref{hs4generators}).} of the) $A(4)$ algebra which is generated by all formal power series of the oscillators
\begin{eqnarray}
\label{hs4generators}
f(y,\bar{y})=\sum_{n,m,n+m=\text{even}}^{\infty}\frac{1}{2i\ m!\ n!}
f^{\alpha_{1}\dots\alpha_{n},\dot{\beta}_{1}\dots\dot{\beta}_{m}}
y_{\alpha_{1}}\dots y_{\alpha_{n}}\bar{y}_{\dot{\beta}_1}\dots\bar{y}_{\dot{\beta}_{m}}\ ,
\end{eqnarray}
with symmetric coefficients $f^{\alpha_{1}\dots\alpha_{n},\dot{\beta}_{1}\dots\dot{\beta}_{m}}=f^{\{\alpha_{1}\dots\alpha_{n}\},\dot{\beta}_{1}\dots\dot{\beta}_{m}}=f^{\alpha_{1}\dots\alpha_{n},\{\dot{\beta}_{1}\dots\dot{\beta}_{m}\}}$. The higher spin Lie algebra $hs(4)$ is the associative algebra $A(4)$ with the commutator as it's Lie bracket. Fields with spin $s$ are then identified with polynomials satisfying $f(ty,t\bar{y})=t^{2(s-1)}f(y,\bar{y})$. For instance the spin 2 sector is spanned by thegenerators of $sp(4)$ defined in (\ref{sp4generators}).

We may define the involutive automorphisms $\pi$ and $\bar{\pi}$ on the $hs(4)$ algebra as follows,
\begin{eqnarray}
\label{ypi}
\pi f(y,\bar{y})= f(-y,\bar{y})\quad\text{and}\quad\bar{\pi} f(y,\bar{y})=f(y,-\bar{y})\ .
\end{eqnarray}
It is evident from (\ref{hs4generators}) that the action of $\pi\bar{\pi}$ on Bosonic fields is the identity while when acting on Fermionic fields, $r_{\alpha}$, we have instead $\pi\bar{\pi}(r_{\alpha})=-r_{\alpha}$. This will prove to be useful when writing down the reality conditions on the physical fields. As was mentioned above, the spin 2 sector of the algebra is spanned by the generators of $sp(4)$, however, as we emphasized in section 2, the unfolded formulation of GR demands a $0$-\emph{form} which would break the $sp(4)$ symmetry down to $sl(2,\mathbb{C})$ so as to identify the frame fields. A symmetry breaking mechanism is provided by a scalar field living in the so called ``\emph{twisted adjoint representation}". This representation make explicit use of the automorphisms $\pi$ and $\bar{\pi}$ which leave the generators $L$ and $\bar{L}$ invariant while the generators $P_{\alpha\dot{\beta}}$ change sign. Furthermore, as we shall prove, the generators $\{L,\bar{L}\}$ do generate the $sl(2,\mathbb{C})\simeq\mathfrak{so}(1,3)$ algebra. It is then convenient to identify the $sl(2,\mathbb{C})$ connection and the frame fields as the $\{L,\bar{L}\}$ and $P_{\alpha\dot{\beta}}$ components of the $sp(4)$ connection, $A$, that is
\begin{eqnarray}
\label{GRconnection}
\omega_{a}&=&\frac{1}{4i}\left(A_{a}^{\ \alpha\beta}L_{\alpha\beta}+\bar{A}_{a}^{\ \dot{\alpha}\dot{\beta}}
\bar{L}_{\dot{\alpha}\dot{\beta}}\right),\\
\label{vierbein}
e_{a}&=&\frac{l}{2i}\left(A_{a}^{\ \alpha\dot{\beta}}P_{\alpha\dot{\beta}}\right)\ .
\end{eqnarray}
Here, $l$ is a length scale related to the cosmological constant via $\Lambda=-\frac{1}{l^{2}}$, and $a=0,1-3$ is the space time index.

\subsection{The Lorentz sector of the Lie algebra}

We shall now prove that the $\{L,\bar{L}\}$ subalgebra of $sp(4)$ is isomorphic to the (complexified) Lorentz algebra. This is best seen by defining $M_{\alpha\beta}$ and $\bar{M}_{\dot{\alpha}\dot{\beta}}$ as
\begin{eqnarray}
\label{LbarLsigma}
M_{\alpha\beta}=J_{ab}\epsilon^{\dot{\gamma}\dot{\delta}}
\sigma^{a}_{\ \alpha\dot{\gamma}}\sigma^{b}_{\ \beta\dot{\delta}}\ \text{ and }
\ \bar{M}_{\dot{\alpha}\dot{\beta}}=J_{ab}\epsilon^{\gamma\delta}
\sigma^{a}_{\ \gamma\dot{\alpha}}\sigma^{b}_{\ \delta\dot{\beta}}\ ,
\end{eqnarray}
where $J_{ab}=J_{[ab]}$ ($a,b=0,1-3$) are the generators of $so(1,3)$. Using the commutation relations for $J_{ab}$,
\begin{eqnarray}
\label{commutation-so(1,3)}
\left[J_{ab},J_{cd}\right]=i\left(\eta_{ad}J_{bc}-\eta_{ac}J_{bd}-\eta_{bd}J_{ac}+
\eta_{bc}J_{ad}\right)\ ,
\end{eqnarray}
and the identity
\begin{eqnarray}
\label{identity}
\eta_{ab}\ \sigma^{a}_{\ \alpha\dot{\alpha}}\sigma^{b}_{\ \beta\dot{\beta}}= 
-2\epsilon_{\alpha\beta}\epsilon_{\dot{\alpha}\dot{\beta}}\ ,
\end{eqnarray}
we may compute the commutation relations for $M$ and $\bar{M}$. We have
\begin{eqnarray}
\nonumber
\left[M_{\alpha\beta},\bar{M}_{\dot{\gamma}\dot{\delta}}\right]&=&
i\epsilon^{\dot{\alpha}\dot{\beta}}
\epsilon^{\gamma\delta}\left(\eta_{ad}J_{bc}-\eta_{ac}J_{bd}-\eta_{bd}J_{ac}+
\eta_{bc}J_{ad}\right)\sigma^{a}_{\ \alpha\dot{\alpha}}\sigma^{b}_{\ \beta\dot{\beta}}
\sigma^{c}_{\ \gamma\dot{\gamma}}\sigma^{d}_{\ \delta\dot{\delta}}
\\ \nonumber
&=& 2iJ_{ab}\Big(
\sigma^{a}_{\ \beta\dot{\delta}}\sigma^{b}_{\ \alpha\dot{\gamma}}+
\sigma^{a}_{\ \beta\dot{\gamma}}\sigma^{b}_{\ \alpha\dot{\delta}}+
\sigma^{a}_{\ \alpha\dot{\delta}}\sigma^{b}_{\ \beta\dot{\gamma}}+
\sigma^{a}_{\ \alpha\dot{\gamma}}\sigma^{b}_{\ \beta\dot{\delta}}
\Big)
\\ \label{MbarM-barMM}
&=&0\ ,
\end{eqnarray}
where we've used the antisymmetry of $J_{ab}$, and
\begin{eqnarray}
\nonumber
\Big[M_{\alpha\beta},M_{\gamma\delta}\Big]&=& i\epsilon^{\dot{\alpha}\dot{\beta}}
\epsilon^{\dot{\gamma}\dot{\delta}}\left(\eta_{ad}J_{bc}-\eta_{ac}J_{bd}-\eta_{bd}J_{ac}+
\eta_{bc}J_{ad}\right)\sigma^{a}_{\ \alpha\dot{\alpha}}\sigma^{b}_{\ \beta\dot{\beta}}
\sigma^{c}_{\ \gamma\dot{\gamma}}\sigma^{d}_{\ \delta\dot{\delta}}\\
\nonumber
&=& i\epsilon^{\dot{\alpha}\dot{\beta}}\epsilon^{\dot{\gamma}\dot{\delta}}\bigg[
\left(\eta_{ad}
\sigma^{a}_{\ \alpha\dot{\alpha}}\sigma^{d}_{\ \delta\dot{\delta}}\right)
J_{bc}\sigma^{b}_{\ \beta\dot{\beta}}\sigma^{c}_{\ \gamma\dot{\gamma}}-
\left(\eta_{ac}
\sigma^{a}_{\ \alpha\dot{\alpha}}\sigma^{c}_{\ \gamma\dot{\gamma}}\right)
J_{bd}\sigma^{b}_{\ \beta\dot{\beta}}\sigma^{d}_{\ \delta\dot{\delta}}
\\ \nonumber
&&-\left(\eta_{bd}
\sigma^{b}_{\ \beta\dot{\beta}}\sigma^{d}_{\ \delta\dot{\delta}}\right)
J_{ac}\sigma^{a}_{\ \alpha\dot{\alpha}}\sigma^{c}_{\ \gamma\dot{\gamma}}+
\left(\eta_{bc}
\sigma^{b}_{\ \beta\dot{\beta}}\sigma^{c}_{\ \gamma\dot{\gamma}}\right)
J_{ad}\sigma^{a}_{\ \alpha\dot{\alpha}}\sigma^{d}_{\ \delta\dot{\delta}}
\bigg]\ 
\\
\nonumber
&=&-2i\epsilon^{\dot{\alpha}\dot{\beta}}\epsilon^{\dot{\gamma}\dot{\delta}}\Big[
\left(\epsilon_{\alpha\delta}\epsilon_{\dot{\alpha}\dot{\delta}}\right)
J_{bc}\sigma^{b}_{\ \beta\dot{\beta}}\sigma^{c}_{\ \gamma\dot{\gamma}}-
\left(\epsilon_{\alpha\gamma}\epsilon_{\dot{\alpha}\dot{\gamma}}\right)
J_{bd}\sigma^{b}_{\ \beta\dot{\beta}}\sigma^{d}_{\ \delta\dot{\delta}}
-\left(\epsilon_{\beta\delta}\epsilon_{\dot{\beta}\dot{\delta}}\right)
J_{ac}\sigma^{a}_{\ \alpha\dot{\alpha}}\sigma^{c}_{\ \gamma\dot{\gamma}}
\\ \nonumber
&&+
\left(\epsilon_{\beta\gamma}\epsilon_{\dot{\beta}\dot{\gamma}}\right)
J_{ad}\sigma^{a}_{\ \alpha\dot{\alpha}}\sigma^{d}_{\ \delta\dot{\delta}}
\Big]\ 
\\ \label{MM-MM}
&=& 2iJ_{ab}\Big(
\epsilon_{\alpha\delta}\epsilon^{\dot{\beta}\dot{\gamma}}
\sigma^{a}_{\ \beta\dot{\beta}}\sigma^{b}_{\ \gamma\dot{\gamma}}+
\epsilon_{\alpha\gamma}\epsilon^{\dot{\beta}\dot{\delta}}
\sigma^{a}_{\ \beta\dot{\beta}}\sigma^{b}_{\ \delta\dot{\delta}}+
\epsilon_{\beta\delta}\epsilon^{\dot{\alpha}\dot{\gamma}}
\sigma^{a}_{\ \alpha\dot{\alpha}}\sigma^{b}_{\ \gamma\dot{\gamma}}+
\epsilon_{\beta\gamma}\epsilon^{\dot{\alpha}\dot{\delta}}
\sigma^{a}_{\ \alpha\dot{\alpha}}\sigma^{b}_{\ \delta\dot{\delta}}
\Big)\ 
\end{eqnarray}
A similar relation holds for the commutation of $\bar{M}$ with itself. Using the definition of $M$ in (\ref{LbarLsigma}), we see that (\ref{MM-MM}) is equal to the \emph{rhs} of (\ref{commutation-sp(4)}). Consequently, $M$ and $\bar{M}$ satisfy the same set of commutation relations as $L$ and $\bar{L}$, and therefore, they generate the same Lie algebra.

\subsection{The supertrace}

For the purpose of writing down an action, we are going to need a notion of trace for the fields valued in $hs(4)$. A suitable choice is the ``\emph{supertrace}'' defined by $\str[f(y,\bar{y})]=f(0,0)$. It is then essential to check if the supertrace reduces to the standard notion of trace in the Lorentz sector of the algebra. Note that the supertrace of the product of two elements, $f(y,\bar{y})$ and $g(y,\bar{y})$, each of which has a formal power series expansion of the form (\ref{hs4generators}) receives contributions from all orders since every term in the product has to be Weyl ordered before performing the supertrace.

To show that the supertrace operation in the $\{L,\bar{L}\}$ subalgebra of $hs(4)$ reduces to the standard trace operation in $so(1,3)$, we'll use the elements $M$ and $\bar{M}$ defined by (\ref{LbarLsigma}). We then claim
\begin{eqnarray}
\label{streqtr1}
\str[L_{\alpha\beta}]&=&-\frac{i}{\sqrt{8}}tr[M_{\alpha\beta}]=0\ ,\\
\label{streqtr2}
\str[L_{\alpha\beta}L_{\gamma\delta}]&=&-\frac{1}{8}tr[M_{\alpha\beta}M_{\gamma\delta}]\ ,
\end{eqnarray}
and similar relations hold for $\bar{L}$ and $\bar{M}$. The first equation is trivial, since $L_{\alpha\beta}=y_{\{\alpha}y_{\beta\}}$ is already Weyl ordered and $J_{ab}$ are traceless; to prove the second equation we first have to calculate all the components of $\str(y_{\alpha}y_{\beta}y_{\gamma}y_{\delta})$ explicitly. Using the commutation relations for the oscillators (\ref{ycommutation}) we have
\begin{eqnarray}
\label{stry1y2y1y2}
\str(y_{1}y_{2}y_{1}y_{2})=\frac{1}{6}\str(y_{1}y_{1}y_{2}y_{2}+y_{1}y_{2}y_{1}y_{2}+
y_{1}y_{2}y_{2}y_{1}+y_{2}y_{1}y_{1}y_{2}+y_{2}y_{1}y_{2}y_{1}+y_{2}y_{2}y_{1}y_{1})=0\ .
\end{eqnarray}
Therefore, the nonzero components of $\str(y_{\alpha}y_{\beta}y_{\gamma}y_{\delta})$ are
\begin{eqnarray}
\label{stry4neq0}
\str(y_{1}y_{2}y_{2}y_{1})=\str(y_{2}y_{1}y_{1}y_{2})=-\str(y_{1}y_{1}y_{2}y_{2})=
-\str(y_{2}y_{2}y_{1}y_{1})=2\ .
\end{eqnarray}
The \emph{lhs} of equation (\ref{streqtr2}) is then
\begin{eqnarray}
\label{strLL}
\str(L_{\alpha\beta}L_{\gamma\delta})=\str(y_{\alpha}y_{\beta}y_{\gamma}y_{\delta})+
\epsilon_{\alpha\beta}\epsilon_{\gamma\delta}\ , 
\end{eqnarray}
or in terms of components,
\begin{eqnarray}
\label{strLL-comp}
\str(L_{11}L_{22})=\str(L_{22}L_{11})=-2 \str(L_{12}L_{12})=-2\ ,
\end{eqnarray}
with all other components, not related by symmetry to the components above, are zero (recall that $L_{\alpha\beta}$ is symmetric). The \emph{rhs} of equation (\ref{streqtr2}) is
\begin{eqnarray}
\nonumber
tr(M_{\alpha\beta}M_{\gamma\delta})&=&tr(J_{ab}J_{cd})\epsilon^{\dot{\mu}\dot{\nu}}
\epsilon^{\dot{\lambda}\dot{\sigma}}\sigma^{a}_{\ \alpha\dot{\mu}}\sigma^{b}_{\ \beta\dot{\nu}}
\sigma^{c}_{\ \gamma\dot{\lambda}}\sigma^{d}_{\ \delta\dot{\sigma}}\\
\label{trMM}
&=&(\eta_{ac}\eta_{bd}-\eta_{ad}\eta_{bc})
(\sigma^{a}_{\ \alpha\dot{1}}\sigma^{b}_{\ \beta\dot{2}}-
\sigma^{a}_{\ \alpha\dot{2}}\sigma^{b}_{\ \beta\dot{1}})
(\sigma^{c}_{\ \gamma\dot{1}}\sigma^{d}_{\ \delta\dot{2}}-
\sigma^{c}_{\ \gamma\dot{2}}\sigma^{d}_{\ \delta\dot{1}})\ .
\end{eqnarray}
Plugging in the explicit form of the Pauli matrices
\begin{eqnarray}
\sigma^{0}=\begin{pmatrix} 1&0\\0&1\end{pmatrix},\ 
\sigma^{1}=\begin{pmatrix} 0&1\\1&0\end{pmatrix},\ 
\sigma^{2}=\begin{pmatrix} 0&-i\\i&0\end{pmatrix},\ 
\sigma^{3}=\begin{pmatrix} 1&0\\0&-1\end{pmatrix},\ 
\end{eqnarray}
we find that the nonzero components of $tr(M_{\alpha\beta}M_{\gamma\delta})$ are (note that $M_{\alpha\beta}$ is symmetric),
\begin{eqnarray}
\label{trMM-comp}
tr(M_{11}M_{22})=tr(M_{22}M_{11})=-2 tr(M_{12}M_{12})=16\ .
\end{eqnarray}
This final result, when compared to (\ref{strLL-comp}), completes the proof of (\ref{streqtr2}). We therefore conclude that the supertrace is the suitable generalization of the trace operation in the lower spin algebra. Before we move on, there is one last property of the supertrace that is of our interest, and that is the invariance under cyclic permutation, i.e.
\begin{eqnarray}
\label{cyclicity}
\str\big(fg\big)=\str\big(g\ \pi\bar{\pi}(f)\big)=\str\big(gf\big)
\end{eqnarray}
where $f$ and $g$ live in the Lie algebra $hs(4)$. Note that in the presence of Fermionic fields we have to use graded cyclic permutation.
\subsection{The twisted adjoint representation}

Before writing down Vasiliev's theory of higher spin gauge fields, we need to introduce the ``twisted adjoint representation" which, as we mentioned earlier, is required for the theory to have non-trivial dynamics.

Let $B$ be a 0-\emph{form} valued in the twisted adjoint representation of $hs(4)$. The twisted adjoint action of $B$ on a generic element of $hs(4)$, $f(Y)$, is defined as follows 
\begin{eqnarray}
\label{twistedadjoinaction}
(f)\tilde{\text{adj}}B\equiv[f,B]_{\pi}=fB-B\pi(f)\ .
\end{eqnarray}
The role of $\Gamma$ in section 2 is then played by a scalar in the twisted adjoint representation.

\section{Vasiliev's theory of higher spin gauge fields}

We may now start putting all the pieces together to write down the simplest (interacting) higher
spin gauge theory. To write down the equations of motion in the unfolded form, we need to introduce
an infinite number of auxiliary fields. To see that this has to be the case, note that to unfold
a set of dynamical equations, the number of required auxiliary fields grows faster, or at least as fast,
as the highest order of derivatives in the equations. In the presence of interacting 
higher spin gauge fields, the order of derivatives grows at least linearly with the highest spin 
present \cite{Vasiliev1996}.

It turns out that a systematic way to add infinitely many auxiliary fields is to extend the 
oscillator expansion of the fields, equation (\ref{hs4generators}), to a new set of 
oscillators $z^{\alpha},\bar{z}^{\dot{\alpha}}$. To write down the constraint equations that 
express the auxiliary fields in terms of the physical fields, we need to introduce a differential 
structure on the $Z$-\emph{space} and to extend the spacetime exterior
algebra to include differential forms on this space. The equations of motion will then be first-order 
differential equations where both the spacetime derivatives and $Z$-space derivatives appear.

\subsection{The $Z$-extension of the Lie algebra}

To introduce all of (the infinite number of) the auxiliary fields in one go, we introduce a new set of 
oscillators $z^{\alpha}$ and $\bar{z}^{\dot{\beta}}$ with the following commutation relations
\begin{eqnarray}
\label{zcommutation}
[z_{\alpha},z_{\beta}]=-2i\epsilon_{\alpha\beta}\ ,\
[\bar{z}_{\dot{\alpha}},\bar{z}_{\dot{\beta}}]=-2i\epsilon_{\dot{\alpha}\dot{\beta}}\ ,\ 
[z_{\alpha},\bar{z}_{\dot{\beta}}]=0\ ,
\end{eqnarray}
and the two set of oscillators $\{y,\bar{y}\}$ and $\{z,\bar{z}\}$ commute. We must now extend
the formal power series (\ref{hs4generators}) to include the new oscillators; the new
formal power series is
\begin{eqnarray}
\label{fields}
f(Y,Z)=\sum_{m,n,p,q}\frac{-1}{4n!m!p!q!}f^{\alpha(n)\dot{\beta}(m)\gamma(p)\dot{\delta}(q)}
Y_{\alpha(n),\dot{\beta}(m)}Z_{\gamma(p),\dot{\delta}(q)}\ ,
\end{eqnarray}
where the restrictions, $m+n=\text{even}$ and $p+q=\text{even}$, are implicit. Here $Y_{\alpha(n),\dot{\beta}(m)}$
is a shorthand for $Y_{\alpha(n),\dot{\beta}(m)}=y_{\alpha_{1}}\dots y_{\alpha_{n}}
\bar{y}_{\dot{\beta}_{1}}\dots\bar{y}_{\dot{\beta}_{m}}$,
where an implicit symmetrization of the indices with the same greek letter is understood.

The action of the automorphisms $\pi$ and $\bar{\pi}$ is extended to include the $Z$ 
oscillators so that
\begin{eqnarray}
\label{automorphism}
\pi\left(f(y,\bar{y},z,\bar{z})\right)=f(-y,\bar{y},-z,\bar{z}),\quad\text{and}\quad
\bar{\pi}\left(f(y,\bar{y},z,\bar{z})\right)=f(y,-\bar{y},z,-\bar{z})\ .
\end{eqnarray}
Note that one can implement the action of these automorphisms by using the Klein 
operators 
\begin{eqnarray}
\label{klein}
\kappa=\text{exp}(iy^{\alpha}z_{\alpha}),\quad\text{and}\quad
\bar{\kappa}=\text{exp}(-i\bar{y}^{\dot{\alpha}}\bar{z}_{\dot{\alpha}}),
\end{eqnarray}
the action of which are given by
\begin{eqnarray}
\label{kleinaction}
f\kappa=\kappa\pi(f),\quad f\bar{\kappa}=\bar{\kappa}\bar{\pi}(f),\quad\text{and}\quad
\kappa^{2}=\bar{\kappa}^{2}=1\ .
\end{eqnarray}

The supertrace operation is also generalized to be $\str(f(Y,Z))=f(0,0)$. We may further
define a partial supertrace $\str_{Z}$ which simply projects the elements 
of the extended algebra, (\ref{fields}), back to the elements of $hs(4)$, (\ref{hs4generators}).

\subsection{Differential structure on the $Z$-space}

The last tool required for writing down Vasiliev's equations of motion is a 
differential structure on the space of $Z$ oscillators. Let us define the 
partial derivatives with respect to $z^{\alpha}$ and $\bar{z}^{\dot{\alpha}}$ as follows,
\begin{eqnarray}
\label{zpartiald}
\frac{\partial}{\partial z^{\alpha}}\ .\ \equiv\frac{i}{2}[z_{\alpha},\ .\ ],\quad
\text{and}\quad\frac{\partial}{\partial \bar{z}^{\dot{\alpha}}}\ .\ \equiv
\frac{i}{2}[\bar{z}_{\dot{\alpha}},\ .\ ]\ .
\end{eqnarray}
The exterior algebra over spacetime may now be extended to include the exterior algebra 
on the $Z$-space with the exterior derivative defined by
\begin{eqnarray}
\label{zexteriord}
d_{Z}\equiv dz^{\alpha}\frac{\partial}{\partial z^{\alpha}}+d\bar{z}^{\dot{\alpha}}
\frac{\partial}{\partial \bar{z}^{\dot{\alpha}}}\ .
\end{eqnarray}
A basis of one forms is then given by the set $\{dx^{a},dz^{\alpha},d\bar{z}^{\dot{\beta}}\}$, 
and all the elements of the set anticommute. The generalized exterior derivative is given by 
$\hat{d}=d+d_{Z}$\ .

Note that for the automorphism $\pi$ and $\bar{\pi}$ act on the basis of one-form in the 
$Z$-space and therefore the components of $p$-forms ($p$=odd) pick up and extra minus 
sign when the automorphism $\pi\bar{\pi}$ is applied. This means that for the components
of these differential forms have the same form as the formal power series (\ref{fields})
except that instead of the restriction $p+q=$even, the restriction $p+q=$odd is imposed
(the restriction $n+m=$even, is unchanged).

\subsection{The connection and curvature}

We are finally ready to write down Vasiliev's equations of motion for higher spin gauge fields. 

As befits a broken topological field theory, the fields consist of a connection $A$ and a symmetry breaking scalar field $B$.  

The connection 1-\emph{form}, $A$ is valued in the extended algebra of formal power series of $Y$'s and $Z$'s.

The  0-\emph{form} $B$ lives in the twisted adjoint representation of the same algebra. 

We may expand the connection 1-\emph{form} in the basis of 1-\emph{forms} as
\begin{eqnarray}
\nonumber
A&=&dx^{a}W_{a}+dz^{\alpha}A_{\alpha}+d\bar{z}^{\dot{\alpha}}\bar{A}_{\dot{\alpha}}\\
\label{Acomponents}
&=&dx^{a}W_{a}+\frac{1}{2i}dz^{\alpha}\left(z_{\alpha}-s_{\alpha}\right)+
\frac{1}{2i}d\bar{z}^{\dot{\alpha}}\left(\bar{z}_{\dot{\alpha}}-\bar{s}_{\dot{\alpha}}\right)\ .
\end{eqnarray}

The generalized covariant exterior derivative is then
\begin{eqnarray}
\nonumber
\hat{D}\ .&=&d\ .\ +d_{Z}\ .\ +[A,\ .\ ]\\
\label{gced}
&=&d\ .\ +[W,\ .\ ]-\frac{1}{2i}[S,\ .\ ]
\end{eqnarray}
where $S=dz^{\alpha}s_{\alpha}+d\bar{z}^{\dot{\alpha}}\bar{s}_{\dot{\alpha}}$, and the contribution of $d_{Z}$ and the $z_{\alpha}$ and $\bar{z}_{\dot{\alpha}}$ terms in $A_{\alpha}$ and $\bar{A}_{\dot{\alpha}}$ cancel. 

The generalized curvature of $A_a$, called $F_{ab}$ is valued in the same algebra as $A_a$ and has the form, 

\begin{eqnarray}
\nonumber
F&=&\hat{d}A+A\wedge A\\
\label{f}
&=&dW+W\wedge W-\frac{1}{2i}(dS+[W,S])-\frac{1}{4}[S,S],
\end{eqnarray}

We will need the covariant derivative of $B$, 
called, $\hat{D}B$.  It  is given by
\begin{eqnarray}
\nonumber
\hat{D}B &=& dB+[W,B]_{\pi}-\frac{1}{2i}[dz^{\alpha}s_{\alpha},B]_{\pi}-
\frac{1}{2i}[d\bar{z}^{\dot{\alpha}}\bar{s}_{\dot{\alpha}},B]_{\bar{\pi}}\\
\label{db}
&=& dB+[W,B]_{\pi}-\frac{1}{2i}dz^{\alpha}\{s_{\alpha},B\}_{\pi}-
\frac{1}{2i}d\bar{z}^{\dot{\alpha}}\{\bar{s}_{\dot{\alpha}},B\}_{\bar{\pi}}\ .
\end{eqnarray}

\subsection{Vasiliev's equations of motion}

We can now finally state Vasiliev's equations of motion.

\begin{eqnarray}
\label{veoma}
F&=&\Sigma,\\
\label{veomb}
\hat{D}B&=&0,
\end{eqnarray}

In the above equations, $\Sigma$ is a two form which is a function of $B$.  It is given by
\begin{eqnarray}
\label{sigma}
\Sigma=dz^{2}\ i(1-B\kappa)+d\bar{z}^{2}\ i(1-B\bar{\kappa}).
\end{eqnarray}
The general form of $\Sigma$ is fixed by demanding Lorentz-covariance\footnote{
See \cite{Iazeolla}, section 5.2.1\ .}.

The equation (\ref{veoma}) constrains the curvature of $A_a$, in a way that is very suggestive of a broken topological field theory.
We will see in the following section how to write an action for it.

\section{An  action principle}

We are almost ready to write down the action corresponding to Vasiliev's field equations (\ref{veoma},\ref{veomb}).  We need some final 
technical preliminaries for dealing with integrals over our infinite component fields. 

\subsection{Technical preliminaries}

To find an action whose extremum is determined by Vasiliev's equations of motion,
we need to construct a spacetime $4$-\emph{form} valued in $\mathbb{R}$. Since all 
physical fields are defined at $Z=0$ (recall that we introduced the $Z$ oscillators 
to add the auxiliary fields to the theory), it seems reasonable to start with an 
$8$-\emph{form} valued in the extended Lie algebra, and to use the 
$Z$-space Hodge star operator $*_{Z}$ to reduce it to a spacetime $4$-\emph{form}
which is also valued in the Lie algebra. We may then use the supertrace 
operation to get the desired $\mathbb{R}$-valued $4$-\emph{form}.
Note that the ``volume form'' in the $Z$-space is given by
\begin{eqnarray}
\label{zvolume}
dZ^{4}=\frac{1}{4}dz^{\alpha}\wedge dz_{\alpha}
\wedge d\bar{z}^{\dot{\beta}}\wedge d\bar{z}_{\dot{\beta}},
\end{eqnarray}
the $Z$-space Hodge star operator is then defined as follows,
\begin{eqnarray}
*_{Z} 1=dZ^{4},\quad
*_{Z}dz^{\alpha}\wedge dz_{\alpha}=d\bar{z}^{\dot{\beta}}\wedge d\bar{z}_{\dot{\beta}},
\quad\text{and}\quad 
*_{Z}^{2}=1.
\end{eqnarray}

To calculate the variation of an action with respect to the components of the fields, we need to have an explicit formulae for the components of the product of two fields. For the purpose of the following calculation we assume, for convenience, that all indices with the same Greek letter are symmetrized, \emph{i.e.} $f^{\alpha_{1}\dots\alpha_{n}}=f^{\{\alpha_{1}\dots\alpha_{n}\}}$.

Consider the product $Y_{\alpha(n)}Y_{\beta(m)}$ ($n\geq m$). We may expand the components of the product as a polynomial of degree $n+m$ as follows,
\begin{eqnarray}
\label{productlaw}
Y_{\alpha(n)}Y_{\beta(m)}=\sum_{k=0}^{k=m}C_{k}^{m}(n)\left(\epsilon_{\alpha\beta}\right)^{k}
Y_{\alpha(n-k)\beta(m-k)}\ ,
\end{eqnarray}
where $\left(\epsilon_{\alpha\beta}\right)^{k}=\prod_{l=1}^{k}
\epsilon_{\alpha_{n+1-l}\beta_{m+1-l}}$\ . Note that $C_{0}^{m}(n)=1$ and 
$C_{1}^{1}(n)=in$, for all $n\geq m$. We may also write
\begin{eqnarray}
\nonumber
Y_{\alpha(n)}Y_{\beta(m)} &=& Y_{\alpha(n)}y_{\beta_{m}}Y_{\beta(m-1)}
\ =\ Y_{\alpha(n)\beta_{m}}Y_{\beta(m-1)}+in Y_{\alpha(n-1)}\epsilon_{\alpha_{n}\beta_{m}}
Y_{\beta(m-1)}\\
\label{recursiveproduct}
&=& Y_{\alpha(n)\beta(m)}+\sum_{k=1}^{m-1}\left[C_{k}^{m-1}(n+1)+in C_{k-1}^{m-1}(n-1)\right]
(\epsilon_{\alpha\beta})^{k}Y_{\alpha(n-k)\beta(m-k)},
\end{eqnarray}
comparing the \emph{rhs} of (\ref{recursiveproduct}) and (\ref{productlaw}),
we deduce the recursion relation
\begin{eqnarray}
\label{recursionrelation}
C_{k}^{m}(n)=C_{k}^{m-1}(n+1)+in C_{k-1}^{m-1}(n-1),
\end{eqnarray}
and in particular,
\begin{eqnarray}
\label{stracecoefficients}
C_{m}^{m}(n)=\frac{i^{m}n!}{(n-m)!}\ .
\end{eqnarray}
The same analysis apply to $\bar{y}$ and the coefficients are exactly the same. For $z$ and
$\bar{z}$, however, there is an extra minus sign contribution from the commutation of $z$'s, 
thus the recursion relation is
\begin{eqnarray}
\label{zrecursionrelation}
C_{\ k}^{\prime\ m}(n)=C_{\ k}^{\prime\ m-1}(n+1)-in C_{\ k-1}^{\prime\ m-1}(n-1),
\end{eqnarray}
and therefore
\begin{eqnarray}
\label{stracecoefficients-z}
C_{\ m}^{\prime\ m}(n)=\frac{(-i)^{m}n!}{(n-m)!}\ .
\end{eqnarray}

Using the above recursion relations, we can now write the supertrace of the product
of any two fields in terms of their components. Consider two general fields $f(Y,Z)$ and
$g(Y,Z)$, each with an expansion of the form (\ref{fields}).
The supertrace of the product of $f$ and $g$ may be written as
\begin{eqnarray}
\nonumber
\str[f(Y,Z)g(Y,Z)]&=&\sum_{n,m,p,q}\frac{i^{n+m}(-i)^{p+q}}{16\ n!m!p!q!}f_{\alpha(n)\dot{\beta}(m)
\gamma(p)\dot{\delta}(q)}g^{\alpha(n)\dot{\beta}(m)\gamma(p)\dot{\delta}(q)}\\
\label{strofproduct}
&=&
\sum_{n,m,p,q}\frac{(-1)^{\frac{n+m+p+q}{2}}}{16\ n!m!p!q!}f_{\alpha(n)\dot{\beta}(m)
\gamma(p)\dot{\delta}(q)}g^{\alpha(n)\dot{\beta}(m)\gamma(p)\dot{\delta}(q)},
\end{eqnarray}
where we've used the fact that $n+m=$even$=p+q$. It is evident from the above equation 
that the supertrace is invariant under cyclic permutation
since at each order, the cyclic permutation generates a factor of $(-1)^{n+m+p+q}=1$.

Note that the above relation does not hold for the product of the components of two
$Z$-space $1$-\emph{forms}, $r_{\alpha^{\prime}}$ and $s_{\beta^{\prime}}$. In this case we have 
$p+q=$odd so that the supertrace of the product is
\begin{eqnarray}
\label{strofproductofs}
\str(r_{\alpha^{\prime}}s_{\beta^{\prime}})&=&(-i)\sum_{n,m,p,q}\frac{(-1)^{\frac{n+m+p+q-1}{2}}}
{16\ n!m!p!q!}r_{\alpha^{\prime}\ \alpha(n)\dot{\beta}(m)
\gamma(p)\dot{\delta}(q)}s_{\beta^{\prime}}^{\ \alpha(n)\dot{\beta}(m)\gamma(p)\dot{\delta}(q)},
\end{eqnarray}
this is why we've inserted a factor of $\frac{1}{i}$ in the $Z$-space part of the connection
1-\emph{form}, (\ref{Acomponents}).

Finally, let $dV=\frac{1}{4!}\epsilon_{abcd} dx^{a}\wedge dx^{b}\wedge dx^{x}\wedge dx^{d}$ .

\subsection{Proposal for an action}

We may now proceed to the construction of an action for the equations of motion (\ref{veoma},\ref{veomb}).  Our basic strategy will be
to introduce lagrange multipliers times these equations of motion.
These will be 
a 6-\emph{form} $Q(Y,Z;x)$ and a 7-\emph{form} $\lambda(Y,Z;x)$.  Their components are defined by
\begin{eqnarray}
\nonumber
Q &=& 
dV\wedge \left(dz^{2}\ \bar{Q}+d\bar{z}^{2}\ Q+
dz^{\alpha}\wedge d\bar{z}^{\dot{\beta}}\ Q_{\alpha\dot{\beta}}\right)
-\frac{1}{2}dx^{a}\wedge dx^{b}\wedge dZ^{4} Q_{ab}\\
\label{Qcomponents}
&&
-\frac{1}{6}\epsilon_{abcd}
dx^{a}\wedge dx^{b}\wedge dx^{c}\wedge
\left(dz^{\alpha}\wedge d\bar{z}^{2}Q^{d}_{\alpha}+
dz^{2}\wedge d\bar{z}^{\dot{\alpha}}\bar{Q}^{\sigma}_{\dot{\alpha}}\right),
\\
\nonumber
\\
\label{lambdacomponents}
\lambda &=& dV\wedge (dz^{\alpha}\wedge d\bar{z}^{2}\ \lambda_{\alpha}+dz^{2}
\wedge d\bar{z}^{\dot{\alpha}}\ \bar{\lambda}_{\dot{\alpha}})-
\frac{1}{6}\epsilon_{abcd}dx^{a}\wedge dx^{b}\wedge dx^{c}\wedge
dZ^{4}\ \lambda^{d}\ .
\end{eqnarray}
We propose an action of the form
\begin{eqnarray}
\label{theaction}
S&=&\int \str\bigg\{*_{Z}\Big[
2Q\wedge\big(F-\Sigma\big)+\lambda\wedge\hat{D}B
\Big]\bigg\}
\end{eqnarray}
where $F$, $\hat{D}B$ and $\Sigma$ are given by (\ref{f}), (\ref{db}) and (\ref{sigma}) respectively. Since $B$ is lives in the twisted adjoint representation of the Lie algebra, we can identify the $sl(2,\mathbb{C})$ part of the spin 2 fields and construct the frame fields (\ref{vierbein}). Given the frame fields, provided that they are non-degenerate, we may identify $dV$ as the volume form and adopt the normalization $\epsilon_{abcd}\epsilon^{abcd}=-4!$ . We then have $dx^{a}\wedge dx^{b}\wedge dx^{c}\wedge dx^{d}=-dV\epsilon^{abcd}$ and the action can be expanded as follows
\begin{eqnarray}
\nonumber
S &=& \int dV\ \str\Bigg\{
\epsilon^{abcd}Q_{ab}\big(\partial_{c}W_{d}+W_{c}W_{d}\big)-i\big(Q^{a\alpha}D_{a}s_{\alpha}+\bar{Q}^{a\dot{\alpha}}D_{a}\bar{s}_{\dot{\alpha}}\big)-Q^{\alpha\dot{\alpha}}\big[s_{\alpha},\bar{s}_{\dot{\alpha}}\big]
\\ \nonumber &&
+\frac{1}{2}Q\epsilon^{\alpha\beta}\Big(\big[s_{\alpha},s_{\beta}\big]+2i\epsilon_{\alpha\beta}(1-B\kappa)\Big)
+\frac{1}{2}\bar{Q}\epsilon^{\dot{\alpha}\dot{\beta}}\Big(\big[\bar{s}_{\dot{\alpha}},\bar{s}_{\dot{\beta}}\big]+2i\epsilon_{\dot{\alpha}\dot{\beta}}(1-B\bar{\kappa})\Big)
\\ \label{HSaction} &&
-\lambda^{a}\big(D_{a}B\big)_{\pi}
+\frac{1}{2i}\Big(\lambda^{\alpha}\big\{s_{\alpha},B\big\}_{\pi}+
\bar{\lambda}^{\dot{\alpha}}\big\{\bar{s}_{\dot{\alpha}},B\big\}_{\bar{\pi}}\Big)
\Bigg\}\ ,
\end{eqnarray}
with $(D_{a}\ .\ )_{\xi}=d\ .\ +[W,\ .\ ]_{\xi}$ denoting the covariant derivative in the suitable representation. By construction, the variation of the above action with respect to the components of $Q$ an $\lambda$ reproduces Vasiliev's equations of motion (\ref{veoma},\ref{veomb}). Varying the action with respect to $W$, $s$, $\bar{s}$ and $B$ we obtain
\begin{eqnarray}
\label{eomw}
\epsilon^{abcd}D_{b}Q_{cd}+
i\big\{s_{\alpha},Q^{a\alpha}\big\}+i\big\{\bar{s}_{\dot{\alpha}},\bar{Q}^{a\dot{\alpha}}\big\}
+\big[\lambda^{a},B\big]_{\pi}&=&0\ ,
\\ \label{eoms}
D_{a}Q^{a\alpha}-i\big\{\bar{s}_{\dot{\alpha}},Q^{\alpha\dot{\alpha}}\big\}+
i\big\{s^{\alpha},Q\big\}-\frac{1}{2}\big\{\lambda^{\alpha},B\big\}_{\pi}&=&0\ ,
\\ \label{eombs}
D_{a}\bar{Q}^{a\dot{\alpha}}+i\big\{s_{\alpha},Q^{\alpha\dot{\alpha}}\big\}+
i\big\{\bar{s}^{\dot{\alpha}},\bar{Q}\big\}-\frac{1}{2}\big\{\bar{\lambda}^{\dot{\alpha}},B\big\}_{\bar{\pi}}&=&0\ ,
\\ \label{eomb}
D_{a}\lambda^{a}-2i\big(Q\kappa+\bar{Q}\bar{\kappa}\big)-\frac{1}{2i}
\big[s_{\alpha},\lambda^{\alpha}\big]-\frac{1}{2i}
\big[\bar{s}_{\dot{\alpha}},\bar{\lambda}^{\dot{\alpha}}\big] &=& 0\ .
\end{eqnarray}
These are differential equations for the Lagrange multipliers $Q$ and $\lambda$. To check if the action that we wrote down is consistent and has solutions we carry out the Hamiltonian analysis of the theory.

\section{The Hamiltonian formulation}

Enormous insight into the dynamical structures of gauge theories is gotten by looking at their Hamiltonian formulation.  This reveals that gauge symmetries
are generated by first class constraints.  We show here that this is true also for Vasiliev's theory.

\subsection{The $3+1$ decomposition}

Consider the action (\ref{HSaction}) on a Lorentzian manifold $\mathcal{M}$ with topology $\mathbb{R} \times M$ where $M$ is a 3 dimensional Euclidean manifold and $\mathbb{R}$ is assumed to be the timelike dimension. We pick a time coordinate $x^{0}=t$ and a 3-volume form, $dv$, on $M$ such that  $dV=dt\wedge dv$. Using these coordinates we can factor the terms with time derivatives in the action (\ref{HSaction}) and find the canonical momenta. The decomposed action looks like
\begin{eqnarray}
\nonumber
S&=&\int dt\int dv\ \str\bigg\{
\epsilon^{ijk}Q_{jk}\big(\partial_{0}W_{i}\big)+\frac{1}{i}Q^{0\alpha}\big(\partial_{0}s_{\alpha}\big)
+\frac{1}{i}\bar{Q}^{0\dot{\alpha}}\big(\partial_{0}\bar{s}_{\dot{\alpha}}\big)
-\lambda^{0}\big(\partial_{0}B\big)
\\ \nonumber &&
+W_{0}\Big(\epsilon^{ijk}D_{i}Q_{jk}-\big\{s_{\alpha},\frac{1}{i}Q^{0\alpha}\big\}
-\big\{\bar{s}_{\dot{\alpha}},\frac{1}{i}\bar{Q}^{0\dot{\alpha}}\big\}+\big[\lambda_{0},B\big]_{\pi}\Big)
+Q_{0i}\epsilon^{ijk}F_{jk}
\\ \nonumber &&
-iQ^{k\alpha}D_{k}s_{\alpha}-i\bar{Q}^{k\dot{\alpha}}D_{k}s_{\dot{\alpha}}-\lambda^{i}\big(D_{i}B\big)_{\pi}
+\frac{i}{2}\lambda^{\alpha}\big\{s_{\alpha},B\big\}_{\pi}+\frac{i}{2}\bar{\lambda}^{\dot{\alpha}}
\big\{\bar{s}_{\dot{\alpha}},B\big\}_{\bar{\pi}}
\\ \label{3+1action} &&
-Q^{\alpha\dot{\alpha}}\big[s_{\alpha},\bar{s}_{\dot{\alpha}}\big]
+Q\Big(\epsilon^{\alpha\beta}s_{\alpha}s_{\beta}+2i\big(1-B\kappa\big)\Big)+\bar{Q}
\Big(\epsilon^{\dot{\alpha}\dot{\beta}}\bar{s}_{\dot{\alpha}}\bar{s}_{\dot{\beta}}
+2i\big(1-B\bar{\kappa}\big)\Big)
\bigg\}\ .
\end{eqnarray}
We can now read off the canonical momenta conjugate to $W_{i}$, $s_{\alpha}$, $\bar{s}_{\dot{\alpha}}$ and $B$, respectively
\begin{eqnarray}
\label{momenta}
\pi^{i}=\epsilon^{ijk}Q_{jk},\ p^{\alpha}=-iQ^{0\alpha},\ \bar{p}^{\dot{\alpha}}=-iQ^{0\dot{\alpha}}, \text{and}\ 
P=-\lambda^{0}\ .
\end{eqnarray}
Inserting these back into (\ref{3+1action}) we find
\begin{eqnarray}
S &=&\int dt\int dv\ \str\bigg\{
\pi^{i}\partial_{0}W_{i}+p^{\alpha}\partial_{0}s_{\alpha}+\bar{p}^{\dot{\alpha}}
\partial_{0}\bar{s}_{\dot{\alpha}}+P\partial_{0}B+W_{0} G+\epsilon^{ijk}Q_{0i}F_{jk}
\\ \nonumber &&
-iQ^{k\alpha}D_{k}s_{\alpha}-i\bar{Q}^{k\dot{\alpha}}D_{k}s_{\dot{\alpha}}-\lambda^{i}\big(D_{i}B\big)_{\pi}
+\frac{i}{2}\lambda^{\alpha}\big\{s_{\alpha},B\big\}_{\pi}+\frac{i}{2}\bar{\lambda}^{\dot{\alpha}}
\big\{\bar{s}_{\dot{\alpha}},B\big\}_{\bar{\pi}}
\\ \label{decmpaction} &&
-Q^{\alpha\dot{\alpha}}\big[s_{\alpha},\bar{s}_{\dot{\alpha}}\big]
+Q\Big(\epsilon^{\alpha\beta}s_{\alpha}s_{\beta}+2i\big(1-B\kappa\big)\Big)+\bar{Q}
\Big(\epsilon^{\dot{\alpha}\dot{\beta}}\bar{s}_{\dot{\alpha}}\bar{s}_{\dot{\beta}}
+2i\big(1-B\bar{\kappa}\big)\Big)
\bigg\}\ ,
\end{eqnarray}
where $G$ is defined below. Since the functions $W_{0},\ Q_{0i},\ Q^{k\alpha},\ \bar{Q}^{k\dot{\alpha}},\ Q^{\alpha\dot{\alpha}},\ Q,\ 
\bar{Q},\ \lambda^{i},\ \lambda^{\alpha}$ and $\bar{\lambda}^{\dot{\alpha}}$ only appear linearly, with no time derivatives, in the action, they can be treated as Lagrange multipliers giving rise to the constraints
\begin{eqnarray}
\label{gauss}
G\equiv
D_{i}\pi^{i}-\big\{s_{\alpha},p^{\alpha}\big\}-\big\{\bar{s}_{\dot{\alpha}},\bar{p}^{\dot{\alpha}}\big\}
-\big[P,B\big]_{\pi}&\approx&0\ ,
\\ \label{Fij}
F_{ij}=\partial_{i}W_{j}-\partial_{j}W_{i}+\big[W_{i},W_{j}\big]&\approx&0\ ,
\\ \label{Ds}
D_{k}s_{\alpha}=\partial_{k}s_{\alpha}+\big[W_{k},s_{\alpha}\big]&\approx&0\ ,
\\ \label{Dbs}
D_{k}\bar{s}_{\dot{\alpha}}=\partial_{k}\bar{s}_{\dot{\alpha}}+\big[W_{k},\bar{s}_{\dot{\alpha}}\big]
&\approx&0\ ,
\\ \label{DB}
\big(D_{i}B\big)_{\pi}=\partial_{i}B+\big[W_{i},B\big]_{\pi}
&\approx&0\ ,
\\ \label{sbs-bss}
\big[s_{\alpha},\bar{s}_{\dot{\alpha}}\big]&\approx&0\ ,
\\ \label{ss-bk}
h_{\alpha\beta}=\big[s_{\alpha},s_{\beta}\big]+2i\epsilon_{\alpha\beta}\big(1-B\kappa\big)&\approx&0\ ,
\\ \label{bsbs-bbk}
\bar{h}_{\dot{\alpha}\dot{\beta}}=\big[\bar{s}_{\dot{\alpha}},\bar{s}_{\dot{\beta}}\big]+2i\epsilon_{\dot{\alpha}\dot{\beta}}
\big(1-B\bar{\kappa}\big)&\approx&0\ ,
\\ \label{sb}
\big\{s_{\alpha},B\big\}_{\pi}&\approx&0\ ,
\\ \label{bsb}
\big\{\bar{s}_{\dot{\alpha}},B\big\}_{\bar{\pi}}&\approx&0\ .
\end{eqnarray}
The first constraint equation is the generalized Gauss's constraint which is common between gauge theories while the rest are just different components of Vasiliev's equations of motion (\ref{veoma},\ref{veomb}). Consequently, the Hamiltonian is just a linear combination of the constraints
\begin{eqnarray}
\nonumber
H=\int dv\ \str\bigg\{-W_{0}G-\epsilon^{ijk}Q_{0i}F_{jk}+iQ^{k\alpha}D_{k}s_{\alpha}+
i\bar{Q}^{k\dot{\alpha}}
D_{k}\bar{s}_{\dot{\alpha}}+\lambda^{i}\big(D_{i}B\big)_{\pi}+Q^{\alpha\dot{\alpha}}
\big[s_{\alpha},\bar{s}_{\dot{\alpha}}\big]
\\ \label{hamiltonian} 
-Q\Big(s_{\alpha}s^{\alpha}+2i\big(1-B\kappa\big)\Big)-\bar{Q}\Big(
\bar{s}_{\dot{\alpha}}\bar{s}^{\dot{\alpha}}+2i\big(1-B\bar{\kappa}\big)\Big)
+\frac{i}{2}\Big(
\lambda^{\alpha}\big\{s_{\alpha},B\big\}_{\pi}+\bar{\lambda}^{\dot{\alpha}}\big\{\bar{s}_{\dot{\alpha}},
B\big\}_{\bar{\pi}}
\Big)\bigg\}
\end{eqnarray}
We can now carry out the Dirac analysis of the constraints by defining the Poisson bracket and forming the constraint algebra.

\subsection{The constraint algebra}

Given the set of canonical fields $W,\ s,\ \bar{s},\ B$ and their conjugate momenta (\ref{momenta}), which are all elements of the higher-spin Lie algebra and have a formal expansion in power series of $Y$'s and $Z$'s, we define the Poisson bracket as follows,
\begin{eqnarray}
\nonumber
&&\Big\{\widetilde{X}_{1}[f],\widetilde{X}_{2}[g]\Big\}=
16\sum_{n,m,p,q}(-1)^{\frac{n+m+p+q}{2}}n!m!p!q!\int dv\ 
\Bigg[
\hspace{128 pt}
\\ \nonumber &&\qquad
\delta_{\substack{n+m,\text{even}\\ p+q,\text{even}}}\bigg(
\frac{\delta\widetilde{X}_{1}[f]}{\delta W_{k}^{\alpha(n)\dot{\beta}(m)\gamma(p)\dot{\delta}(q)}}\ 
\frac{\delta\widetilde{X}_{2}[g]}{\delta\pi^{k}_{\alpha(n)\dot{\beta}(m)\gamma(p)\dot{\delta}(q)}}
-
\frac{\delta\widetilde{X}_{1}[f]}{\delta\pi^{k}_{\alpha(n)\dot{\beta}(m)\gamma(p)\dot{\delta}(q)}}\ 
\frac{\delta\widetilde{X}_{2}[g]}{\delta W_{k}^{\alpha(n)\dot{\beta}(m)\gamma(p)\dot{\delta}(q)}}
\\ \nonumber &&\qquad
+
\frac{\delta\widetilde{X}_{1}[f]}{\delta B_{\alpha(n)\dot{\beta}(m)\gamma(p)\dot{\delta}(q)}}\ 
\frac{\delta\widetilde{X}_{2}[g]}{\delta P^{\alpha(n)\dot{\beta}(m)\gamma(p)\dot{\delta}(q)}}
-
\frac{\delta\widetilde{X}_{1}[f]}{\delta P^{\alpha(n)\dot{\beta}(m)\gamma(p)\dot{\delta}(q)}}\ 
\frac{\delta\widetilde{X}_{2}[g]}{\delta B_{\alpha(n)\dot{\beta}(m)\gamma(p)\dot{\delta}(q)}}
\bigg)
\\ \nonumber &&\qquad
+\delta_{\substack{n+m,\text{even}\\ p+q,\text{odd}}}\bigg(
\frac{\delta\widetilde{X}_{1}[f]}{\delta s_{\sigma}^{\alpha(n)\dot{\beta}(m)\gamma(p)\dot{\delta}(q)}}\ 
\frac{\delta\widetilde{X}_{2}[g]}{\delta p^{\sigma}_{\alpha(n)\dot{\beta}(m)\gamma(p)\dot{\delta}(q)}}
-\frac{\delta\widetilde{X}_{1}[f]}{\delta p^{\sigma}_{\alpha(n)\dot{\beta}(m)\gamma(p)\dot{\delta}(q)}}\ 
\frac{\delta\widetilde{X}_{2}[g]}{\delta s_{\sigma}^{\alpha(n)\dot{\beta}(m)\gamma(p)\dot{\delta}(q)}}
\\ \label{Poisson} &&\qquad
+\frac{\delta\widetilde{X}_{1}[f]}{\delta \bar{s}_{\dot{\sigma}}^{\alpha(n)\dot{\beta}(m)\gamma(p)\dot{\delta}(q)}}\ 
\frac{\delta\widetilde{X}_{2}[g]}{\delta \bar{p}^{\dot{\sigma}}_{\alpha(n)\dot{\beta}(m)\gamma(p)\dot{\delta}(q)}}
-\frac{\delta\widetilde{X}_{1}[f]}{\delta \bar{p}^{\dot{\sigma}}_{\alpha(n)\dot{\beta}(m)\gamma(p)\dot{\delta}(q)}}\ 
\frac{\delta\widetilde{X}_{2}[g]}{\delta \bar{s}_{\dot{\sigma}}^{\alpha(n)\dot{\beta}(m)\gamma(p)\dot{\delta}(q)}}
\bigg)
\Bigg]
\end{eqnarray}
where we have \emph{smeared out} $X_{1}$ and $X_{2}$ by the elements $f$ and $g$. The general form of a field $X$ smeared out with an element $f$ is
\begin{eqnarray}
\label{smear}
\widetilde{X}[f]=\int dv\ \str\Big\{f\ X\Big\}\  .
\end{eqnarray}
One may verify the following Poisson brackets of the fields and their conjugate momenta
\begin{eqnarray}
\label{poissonwpi}
\Big\{\widetilde{W}_{i}[f^{i}],\widetilde{\pi}^{j}[g_{j}]\Big\}&=&\widetilde{g}_{i}[f^{i}]\ ,
\\ \label{poissonsp}
\Big\{\widetilde{s}_{\alpha}[f^{\alpha}],\widetilde{p}^{\beta}[g_{\beta}]\Big\}&=&
-\widetilde{g}_{\alpha}[f^{\alpha}]\ ,
\\ \label{poissonbsbp}
\Big\{\widetilde{\bar{s}}_{\dot{\alpha}}[f^{\dot{\alpha}}],\widetilde{\bar{p}}^{\dot{\beta}}[g_{\dot{\beta}}]\Big\}&=&
-\widetilde{g}_{\dot{\alpha}}[f^{\dot{\alpha}}]\ ,
\\ \label{poissonbp}
\Big\{\widetilde{B}[f],\widetilde{P}[g]\Big\}&=&\widetilde{g}[\pi(f)]\ .
\end{eqnarray}
Since the momenta appear only in the Gauss's constraint (\ref{gauss}), the Poisson bracket of any pair of the constraints (\ref{Ds} - \ref{bsb}) vanishes. Noting that the Hamiltonian is a linear combination of the constraints, to identify the first and second class constraints, it is sufficient to compute the Poisson bracket of all the constraints with $G$. The resulting Poisson brackets read
\begin{eqnarray}
\label{poissongg}
\Big\{\widetilde{G}[f],\widetilde{G}[g]\Big\}=\widetilde{G}\Big[\big[f,g\big]\Big]&\approx& 0\ ,
\\ \label{poissson2}
\Big\{\widetilde{G}[f],\widetilde{F}_{ij}[g^{ij}]\Big\}=\widetilde{F}_{ij}\Big[\big[f,g^{ij}\big]\Big]&\approx& 0\ ,
\\ \label{poissson3}
\Big\{\widetilde{G}[f],\widetilde{D_{k}s_{\alpha}}[g^{k\alpha}]\Big\}=\widetilde{D_{k}s_{\alpha}}\Big[\big[f,g^{k\alpha}\big]\Big]&\approx& 0\ ,
\\ \label{poissson4}
\Big\{\widetilde{G}[f],\widetilde{D_{k}\bar{s}_{\dot{\alpha}}}[g^{k\dot{\alpha}}]\Big\}=\widetilde{D_{k}\bar{s}_{\dot{\alpha}}}\Big[\big[f,g^{k\dot{\alpha}}\big]\Big]&\approx& 0\ ,
\\ \label{poissson5}
\Big\{\widetilde{G}[f],\widetilde{\big(D_{i}B\big)_{\pi}}[g^{i}]\Big\}=\widetilde{\big(D_{i}B\big)_{\pi}}\Big[\big[f,g^{i}\big]\Big]&\approx& 0\ ,
\\ \label{poissson6}
\Big\{\widetilde{G}[f],\widetilde{\big[s_{\alpha},\bar{s}_{\dot{\alpha}}\big]}[g^{\alpha\dot{\alpha}}]\Big\}=\widetilde{\big[s_{\alpha},\bar{s}_{\dot{\alpha}}\big]}\Big[\big[f,g^{\alpha\dot{\alpha}}\big]\Big]&\approx& 0\ ,
\\ \label{poissson7}
\Big\{\widetilde{G}[f],\widetilde{h}_{\alpha\beta}[g^{\alpha\beta}]\Big\}=\widetilde{h}_{\alpha\beta}\Big[\big[f,g^{\alpha\beta}\big]\Big]&\approx& 0\ ,
\\ \label{poissson8}
\Big\{\widetilde{G}[f],\widetilde{\bar{h}}_{\dot{\alpha}\dot{\beta}}[g^{\dot{\alpha}\dot{\beta}}]\Big\}=\widetilde{\bar{h}}_{\dot{\alpha}\dot{\beta}}\Big[\big[f,g^{\dot{\alpha}\dot{\beta}}\big]\Big]&\approx& 0\ ,
\\ \label{poissson9}
\Big\{\widetilde{G}[f],\widetilde{\big\{s_{\alpha},B\big\}_{\pi}}[g^{\alpha}]\Big\}=\widetilde{\big\{s_{\alpha},B\big\}_{\pi}}\Big[\big[f,g^{\alpha}\big]\Big]&\approx& 0\ ,
\\ \label{poissson10}
\Big\{\widetilde{G}[f],\widetilde{\big\{\bar{s}_{\dot{\alpha}},B\big\}_{\bar{\pi}}}[g^{\dot{\alpha}}]\Big\}=\widetilde{\big\{\bar{s}_{\dot{\alpha}},B\big\}_{\bar{\pi}}}\Big[\big[f,g^{\dot{\alpha}}\big]\Big]&\approx& 0\ .
\end{eqnarray}
It follows that all of the constraints (\ref{gauss}-\ref{bsb}) are first-class and therefore the Hamiltonian (\ref{hamiltonian}) is the first-class Hamiltonian.

It is an easy exercise to check that the Gauss's constraint (\ref{gauss}) generates gauge transformations,
\begin{eqnarray}
\label{pgaussw}
\Big\{\widetilde{G}[\epsilon],\widetilde{W_{i}}[f^{i}]\Big\}&=&\widetilde{D_{i}\epsilon}[f^{i}]\ ,
\\ \label{pgausss}
\Big\{\widetilde{G}[\epsilon],\widetilde{s_{\alpha}}[f^{\alpha}]\Big\}&=&
\widetilde{\big[s_{\alpha},\epsilon\big]}[f^{\alpha}]\ ,
\\ \label{pgaussbs}
\Big\{\widetilde{G}[\epsilon],\widetilde{\bar{s}_{\dot{\alpha}}}[f^{\dot{\alpha}}]\Big\}&=&
\widetilde{\big[\bar{s}_{\dot{\alpha}},\epsilon\big]}[f^{\dot{\alpha}}] \ ,
\\ \label{pgaussb}
\Big\{\widetilde{G}[\epsilon],\widetilde{B}[f]\Big\}&=&\widetilde{\big[\epsilon,B\big]_{\pi}}[-f] \ .
\end{eqnarray}
As a final remark, note that for a gauge transformation with the gauge parameter $\epsilon=\xi^{i}W_{i}$, with $\xi^{i}$ is a real valued vector, equation (\ref{pgaussw}) becomes
\begin{eqnarray}
\nonumber
\Big\{\widetilde{G}[\xi^{j}W_{j}],\widetilde{W_{i}}[f^{i}]\Big\}&=&
\widetilde{D_{i}\big(\xi^{j}W_{j}\big)}[f^{i}]
\\ \nonumber
&=& \widetilde{\big(\mathcal{L}_{\xi}W_{i}+\xi^{j}F_{ij}\big)}[f^{i}]
\\ \label{diffeomorphisms}
&\approx& \widetilde{\big(\mathcal{L}_{\xi}W_{i}\big)}[f^{i}]\ ,
\end{eqnarray}
This is just a (spacial) diffeomorphism, provided that the constraint (\ref{Fij}) is satisfied. A similar relation holds for $s$, $\bar{s}$ and $B$.

\section{Conclusion}

To summarize our results, we've proposed the action (\ref{theaction}) for 4 dimensional massless Bosonic higher spin gauge theory which encodes Vasiliev's equations of motion (\ref{veoma}, \ref{veomb}). Moreover, we've carried out the constraint analysis as a consistency check, and derived the constraint algebra of the theory. The Hamiltonian (\ref{hamiltonian}) is shown to be a linear combination of the constraints (\ref{gauss} - \ref{bsb}) which are all shown to be first class. 

We've further shown that as in Yang-Mills gauge theories and Einstein's gravity, the generalized Gauss's constrain (\ref{gauss}) generates gauge transformations and, on the constraint hypersurface, spacial diffeomorphisms.

\section*{Acknowledgments}
We would like to thank Jaume Gomis for introducing us to subject of higher spin gauge theory and for many insightful comments and discussions. Research at Perimeter Institute for Theoretical Physics is supported in part by the Government of Canada through NSERC and by the Province of Ontario through MRI. One of the authors (N.D.) is also supported in part by NSERC Discovery Grant.

\end{document}